\documentclass{optica-article}

\journal{opticajournal}

\usepackage{lineno}
\usepackage{graphicx}
\usepackage{amsmath, amsfonts, amscd, physics}
\usepackage{setspace}
\usepackage{inputenc}
\usepackage{mathrsfs, dsfont}
\usepackage{mathtools}
\usepackage{stmaryrd}
\usepackage{array}
\usepackage{bm}
\usepackage[T1]{fontenc}
\begin{document}

\title{Two-photon Hong-Ou-Mandel interference and quantum entanglement between the frequency-converted idler photon and the signal photon}

\author{Jiaxuan Wang,\authormark{1,2,*} Alexei V. Sokolov\authormark{1,2}, and Girish S. Agarwal,\authormark{1,2,3}}

\address{\authormark{1}Department of Physics and Astronomy, Texas A\&M University, College Station, Texas 77843, USA\\
\authormark{2}Institute for Quantum Science and Engineering, Texas A\&M University, College Station, Texas 77843, USA\\
\authormark{3}Department of Biological and Agricultural Engineering, Texas A\&M University, College Station, Texas 77843, USA}

\email{\authormark{*}jxnwang@tamu.edu} %% email address is required; see note below about the corresponding author designation

% use {asbstract*} to suppress the copyright line. Copyright information will be added in production

\begin{abstract*} 
Quantum frequency up-conversion is a cutting-edge technique that leverages the interaction between
photons and quantum systems to shift the frequency of single photons from a lower frequency to a higher
frequency. If the photon before up-conversion was one of the entangled pair, then it is important to
understand how much entanglement is preserved after up-conversion. In this study, we present a
theoretical analysis of the transformation of the time-dependent second-order quantum correlations in
photon pairs and find the preservation of such correlations under fairly general conditions. We also
analyze the two-photon Hong-Ou-Mandel interference between the frequency-converted idler photon and
the signal photon. The visibility of the two-photon interference is sensitive to the magnitude of the
frequency conversion, and it improves when the frequency separation between two photons goes down.

\end{abstract*}

%%%%%%%%%%%%%%%%%%%%%%%%%%  body  %%%%%%%%%%%%%%%%%%%%%%%%%%
\section{Introduction}
Optical frequency up-conversion has been the subject of much interest and investigation since its discovery in 1967 by Midwinter and Warner \cite{midwinter1967up}. 
The driving factor behind the importance of up-conversion \cite{ma2012single} has been the possibility of detecting infrared radiation \cite{thew2006low, vandevender2004high, albota2004efficient, langrock2005highly, ma2009experimental} by converting it to
the optical domain as the detection technology in the optical domain is much better.
Thus, techniques for
improving the efficiency of the up-conversion process have been developed \cite{huang2013quantum, PhysRevA.91.013837}.
The improved efficiency is important for many applications, for example, in imaging \cite{barh2019parametric, ashik2019mid}. Of particular importance in the context of quantum
technologies is the up-conversion of a single photon.
The quantum theory of the up-conversion process
has been developed \cite{PhysRev.124.1646, tucker1969quantum}. With the growing interest in
quantum entanglement, Kumar and coworkers \cite{kumar1990quantum, huang1992observation, huang2013quantum} demonstrated preservation of the quantum
correlations between up-converted idler photons and signal photons produced by an optical parametric
amplifier. In particular, they showed the survival of the non-classical intensity correlation between two
photons at 1064 nm when one of these was converted to 532 nm. Since these early experiments, the interest
shifted to single photons, and several experiments studied the quality of the up-converted single photons \cite{PhysRevLett.105.093604, ates2012two}
and verified the preservation of quantum correlations during the process \cite{fan2016integrated, chen2021single, zhu2022spectral}. As known, a versatile
source of single photons is the spontaneous parametric down-conversion process where one has entangled
pairs, i.e., signal and idler photons whose frequencies can be quite different, still exhibiting strong second-order quantum correlations represented by $g^{(2)}(\tau)$. Direct measurement of $g^{(2)}(\tau)$ is difficult as the
correlation time could be 100 fs or less, and one uses the second-order Hong-Ou-Mandel interference \cite{hong1987measurement, barbieri2017hong, walborn2003multimode, jin2018quantum, fabre2021parameter, douce2013direct, boucher2015toolbox, liu2022probing} for getting information at these time scales. In a recent work \cite{tyumenev2022tunable, sokolov2022giving} the correlation time was about 80 ps \cite{hammer2020broadly}, and the frequency of the idler photon
was up-converted by about 120 THz. Tyumenev et al \cite{tyumenev2022tunable} demonstrated the preservation of such correlation
time between the up-converted idler photon and the signal photon. Given the considerable interest in
the quantum properties of the up-converted photons, we present a first principle theoretical calculation of
the non-classical time correlations of the up-converted and down-converted photon and the signal
photon. We find the preservation of such correlations under fairly general conditions. We also analyze the
two-photon Hong-Ou-Mandel interference between the frequency-converted idler photon and the signal
photon. The visibility of the two-photon interference is sensitive to the magnitude of the frequency
conversion, and it improves when the frequency separation between two photons goes down. Our theoretical
results on HOM interference are quite relevant to several experiments, for example, \cite{fan2016integrated, chen2021single, zhu2022spectral} on the quality of frequency-converted single photons.
Our results also apply to cases when the frequency is down-converted though we focus on up-conversion processes.

\section{Second-order Correlation of the Up-shifted Idler and Original Signal Biphotons}

As shown in Fig.~1, we start with the entangled biphoton state \cite{shih2020introduction} comprised of a idler photon of frequency $\omega_{s}$ and an signal photon of frequency $\omega_{i}$
\begin{align}
\ket{\Phi}=\int\int d\omega_{s}d\omega_{i}f(\omega_{s},\omega_{i})a_{s}^{\dagger}(\omega_{s})a_{i}^{\dagger}(\omega_{i})\ket{0,0},
\end{align}
where $a_{s(i)}^{\dagger}(\omega_{s(i)})$ is the creation operator for the idler (signal) mode that satisfies the  $[a_{s(i)}(\omega),a^{\dagger}_{s(i)}(\omega') ] = \delta(\omega - \omega')$, and $[a_{s(i)}(\omega),a_{i(s)}(\omega') ] = [a_{s(i)}(\omega),a^{\dagger}_{i(s)}(\omega') ] = 0$. Function $f(\omega_{s},\omega_{i})$ is the two-mode frequency distribution. We consider the input photon statistics to be an entangled Gaussian distribution
\begin{align}\label{distribution}
f(\omega_{s},\omega_{i})=\frac{1}{\sqrt{2\pi\sigma_{P}\sigma_{-}}}e^{-(\omega_{s}+\omega_{i}-\omega_{p})^{2}/(16\sigma_{P}^{2})}e^{-(\omega_{s}-\omega_{i}-\triangle)^{2}/(4\sigma_{-}^{2})},
\end{align}
where $\omega_{p}$ and $\sigma_{P}$ are the frequency and bandwidth of the pump; $\triangle$ is the frequency difference between the central frequency $\omega_{i0}$ and $\omega_{s0}$; $\sigma_{-}$ is the bandwidth of the photon pairs. The distribution is normalized, i.e.,  $\int\int|f(\omega_{s},\omega_{i})|^2 d\omega_{s} d \omega_{i} = 1$. 
\begin{figure}[tb]
\centering
\includegraphics[width=10cm]{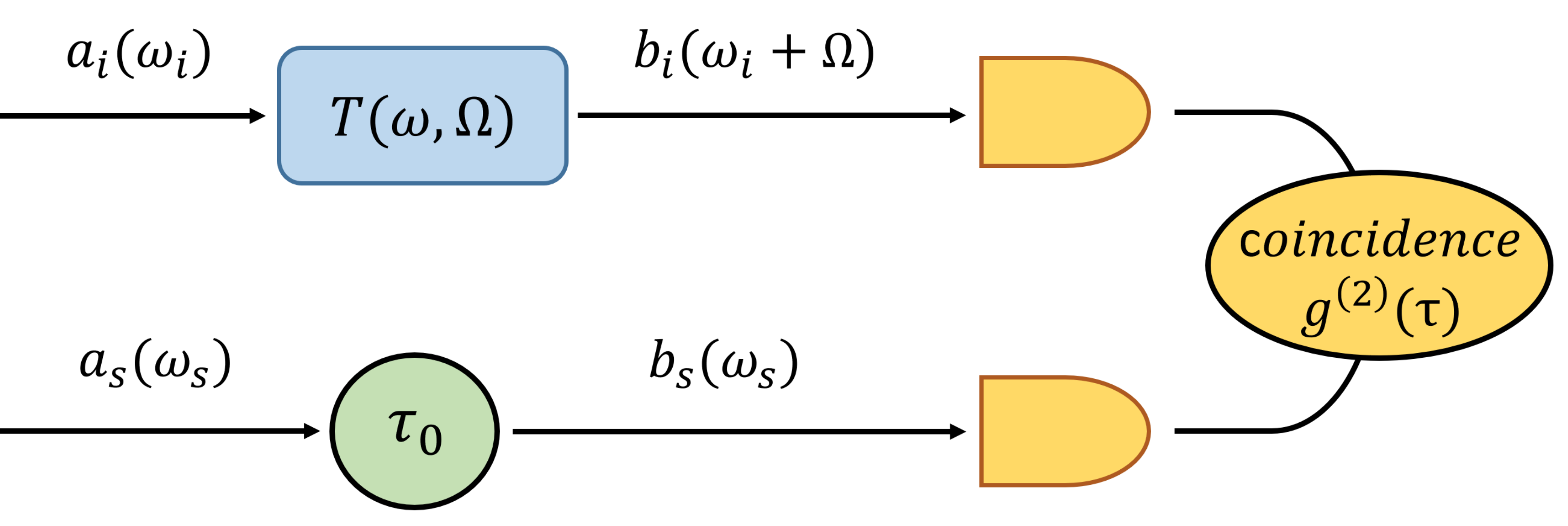}
\caption{Experimental setup for the detection of second-order correlation $g^{(2)}(\tau)$ of the up-shifted idler and original signal biphotons. }
\label{Fig2}
\end{figure}
As for the frequency up-conversion process, The idler photon travels through a medium that up-shifts its frequency
\begin{align} 
b_{i}(\omega)=T(\omega,\Omega)a_{i}(\omega-\Omega)+\eta(\omega,\Omega)a_{v}(\omega),
\end{align}
where $T(\omega,\Omega)$ is the up-conversion rate; $a_{v}(\omega)$ is the vacuum noise creation operator and $\eta(\omega,\Omega)=\sqrt{1-|T(\omega,\Omega)|^2}$. 
The operators $b_i(\omega)$ and $b^{\dagger}_i(\omega')$ satisfy the commutation relations $[b_i(\omega),b^{\dagger}_i(\omega') ] = \delta(\omega - \omega')$. 
This formula has the same format as a transmission output-input relationship in an ordinary sample, while now the only difference is that the output frequency is up-converted, and the conversion rate should be dependent on both the original frequency and the frequency shift. 
Many experiments on frequency conversion use modulation at frequency $\Omega$. Thus the input photons at $\omega$ would be converted to $\omega + \Omega$ and $\omega - \Omega$. The modulation was in the GHz domain for some of the experiments, as noted in the introduction. 
The signal photon frequency remains the same. A phase shift is introduced due to the different paths taken by the photon pair
\begin{align}
b_{s}(\omega)=a_{s}(\omega)e^{-i\omega\tau_0}.
\end{align}
The second order correlation $g^{(2)}(t,t+\tau)$ is defined as
%\begin{figure}[tb]
%\includegraphics[width=7cm]{up_convert.pdf}
%\caption{Scheme of optical frequency up-conversion. The input beam of frequency $\omega$ undergoes a frequency transformation to $\omega+\Omega$ as it passes through the up-conversion sample pumped by a strong beam of frequency $\Omega$. $I_{out}(\omega+\Omega)=T^2I_{in}(\omega)$. There can be a seed of $\omega+\Omega$ from the input. In our study, it is taken to be a vacuum mode.}
%\end{figure}
%\label{Fig1}
\begin{align}
g^{(2)}(t,t+\tau)=\expval{b^{\dagger}_{i}(t)b^{\dagger}_{s}(t+\tau)b_{s}(t+\tau)b_{i}(t)}.
\end{align}
Considering Eq. (1), (3), (4), we obtain
\begin{align}
b_{s}(t+\tau)b_{i}(t)\ket{\Phi}
&=\frac{1}{2\pi}\int\int d\omega_{1}d\omega_{2}T(\omega_{1},\Omega)a_{s}(\omega_{2})a_{i}(\omega_{1}-\Omega)e^{i\omega_{1}t}e^{i\omega_{2}(t+\tau-\tau_{0})} \notag\\
&\times \int\int d\omega_{s}d\omega_{i}f(\omega_{s},\omega_{i})a_{s}^{\dagger}(\omega_{s})a_{i}^{\dagger}(\omega_{i})\ket{0,0}, 
\end{align}
which gives the result for Eq. (5) as
\begin{align}
g^{(2)}(t,t+\tau)&=\frac{1}{4\pi^2}|\int\int d\omega_{1}d\omega_{2}T(\omega_{1},\Omega)\times e^{i\omega_{1}t}e^{i\omega_{2}(t+\tau-\tau_{0})}f(\omega_{1}-\Omega,\omega_{2})|^2. 
\end{align}
Note that the vacuum noise term does not contribute to the normal-ordered correlation. Assuming that the phase-matching is satisfied over the entire frequency width of the idler photon, then the up-conversion rate is a constant $T(\omega,\Omega)=T$ in the range of interest. Applying the parameter transformation $\Omega_{1}=\omega_{1}-\Omega$, Eq. (7) can be simplified to
\begin{align}
g^{(2)}(t,t+\tau)&=\frac{T^2}{4\pi^2}|\int\int d\Omega_{1}d\omega_{2}e^{i\Omega_{1}t}e^{i\omega_{2}(t+\tau-\tau_{0})}f(\Omega_{1},\omega_{2})|^2.
\end{align}
Considering a photon pair following Gaussian distribution as in Eq. (2), we obtain
\begin{align}
g^{(2)}(t,t+\tau)&=\frac{T^{2}}{(2\pi)^{3}\sigma_{P}\sigma_{-}}|\int\int d\Omega_{1}d\omega_{2}e^{i(\omega_{2}+\Omega_{1})\frac{(t+\tau'-\tau_{0})+t}{2}} \notag\\
&\times e^{i(\omega_{2}-\Omega_{1})\frac{(t+\tau-\tau_{0})-t}{2}}e^{-\frac{(\omega_{2}+\Omega_{1}-\omega_{p})^{2}}{16\sigma_{P}^{2}}}e^{-\frac{(\omega_{2}-\Omega_{1}-\triangle)^{2}}{4\sigma_{-}^{2}}}|^{2}. 
\end{align}
By orthogonally transforming parameters to $u=\frac{\omega_{2}+\Omega_{1}}{\sqrt{2}}$ and $v=\frac{\omega_{2}-\Omega_{1}}{\sqrt{2}}$, the double integral reduces to a product of two single-parameter integrals.
\begin{align}
g^{(2)}(t,t+\tau)&=\frac{T^{2}}{(2\pi)^{3}\sigma_{P}\sigma_{-}}|\int due^{iu\frac{2t+\tau-\tau_0}{\sqrt{2}}}e^{-(u-\frac{\omega_{p}}{\sqrt{2}})^{2}/(8\sigma_{P}^{2})}\int dve^{iv\frac{\tau-\tau_0}{\sqrt{2}}}e^{-(v-\frac{\triangle}{\sqrt{2}})^{2}/(2\sigma_{-}^{2})}|^{2}. \notag\\
\end{align}
Notably, the two integrals present in the expression are the Fourier transforms of Gaussian functions. The shifts $\frac{\omega_{p}}{\sqrt{2}}$ and $\frac{\triangle}{\sqrt{2}}$ thus don't contribute to the modulus square. We then obtain
\begin{align}
g^{(2)}(t,t+\tau)=\frac{2\sigma_{P}\sigma_{-}T^{2}}{\pi}e^{-2\sigma_{P}^{2}(2t+\tau-\tau_0)^{2}}e^{-\frac{1}{2}\sigma_{-}^{2}(\tau-\tau_0)^{2}}.
\end{align}
Note that $\omega_{p}$ and $\triangle$ do not explicitly contribute to the formula. This observation is due to the fact that the photon pair in question follows a standard bivariate normal distribution $f_{XY}(x,y)$. The correlation between the two variables, denoted by $\rho$, is described by the given formula \cite{kenney1939mayhematics, whittaker1924calculus}
\begin{align}
f_{XY}(x,y)=\frac{1}{2\pi\sqrt{1-\rho^{2}}}e^{-\frac{x^{2}-2\rho xy+y^{2}}{2(1-\rho^{2})}}.
\end{align}
In our case the correlation $\frac{4\sigma_{p}^{2}-\sigma_{-}^{2}}{4\sigma_{p}^{2}+\sigma_{-}^{2}}$ for the photon statistics $|f(\omega_{s},\omega_{i})|^2$ is a function of $\sigma_{-}$ and $\sigma_{p}$ only. 
\begin{figure}[tb]
\centering
\includegraphics[width=7cm]{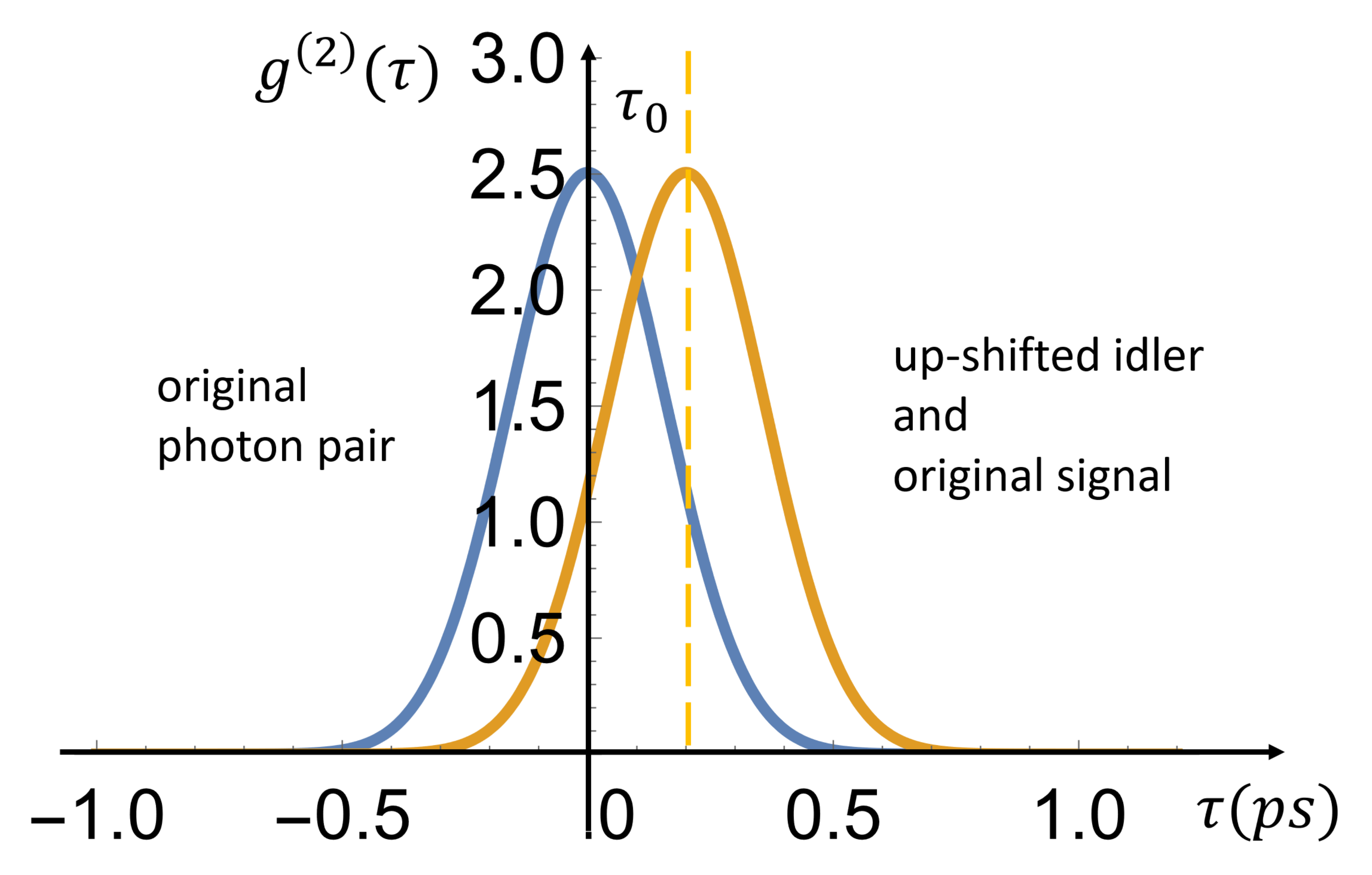}
\caption{Normalized second-order correlation $g^{(2)}(\tau)$ in Eq. (14) of the up-shifted idler and original signal photon pair (yellow), compared to $g^{(2)}(\tau)$ of the original photon pair (blue), for $\sigma_{-} = 2\pi\times 1\:THz$, $\sigma_{p} = \sigma_{-}/10$, $\tau_0 = 0.2\:ps$ and $T_R=100\:ps$.}
\label{Fig3}
\end{figure}

The two-time correlation in Eq. (11) is to be averaged over the resolution time $T_R$ of the detector.
\begin{align} 
g^{(2)}(\tau)=\int_{-T_{R}/2}^{+T_{R}/2}dt g_{2}(t,t+\tau),
\end{align}
which gives 
\begin{align} 
g^{(2)}(\tau)=\frac{1}{2\sqrt{2\pi}}\sigma_{-}T^{2}e^{-\frac{1}{2}\sigma_{-}^{2}(\tau-\tau_0)^{2}}[erf(\sqrt{2}\sigma_{P}(\tau-\tau_0+T_{R}))-erf(\sqrt{2}\sigma_{P}(\tau-\tau_0-T_{R}))], \notag\\
\end{align}
where $erf(x)$ stands for the error function. We consider two scenarios. In the first scenario, when the large-detection time condition is satisfied, i.e.,  $\sigma_{P}T_{R}\rightarrow\infty$ and $T_{R}\gg\tau-\tau_0$, say with $T_R\sim10\:ps$ or $100\:ps$, we obtain
\begin{align}
g^{(2)}(\tau)=\frac{\sigma_{-}T^{2}}{\sqrt{2\pi}}e^{-\frac{1}{2}\sigma_{-}^{2}(\tau-\tau_0)^{2}}.
\end{align}
The FWHM of $g^{(2)}(\tau)$ between the up-shifted idler and original signal photon pair is the same as that between the original photon pairs. This is illustrated in Fig. 2, where the normalized $g^{(2)}(\tau)$ (by dropping the factor $T^2$) is plotted for the up-shifted idler and original signal photon pair, and demonstrates the presence of only a peak shift arising from the difference in path length. 
Note that Fig. 2 is obtained by using the exact result in Eq. (14).
In the second scenario, $\sigma_{P}T_{R}\rightarrow0$, and $\sigma_{P}(\tau-\tau_0)\rightarrow0$ we obtain
\begin{align}
g^{(2)}(\tau)=\frac{T_{R}\sigma_{P}\sigma_{-}T^{2}}{\sqrt{\pi}}e^{-\frac{1}{2}\sigma_{-}^{2}(\tau-\tau_0)^{2}}.
\end{align}
%regarding to $\sigma_-=1/80 \,ps$ and $T_R=50\,ps$ in the experiment \cite{tyumenev2022tunable}. 
In both limits, the width of the $g^{(2)}(\tau)$ function is unchanged except for the phase difference imparted by the up-conversion medium from the correlation when there is no up-conversion (and no $\tau_0$).
\begin{align}
g^{(2)}_0(\tau)\propto\frac{\sigma_{-}}{\sqrt{2\pi}}e^{-\frac{1}{2}\sigma_{-}^{2}(\tau)^{2}}.
\end{align}

Next, we examine the effect of phase-matching on the two-photon correlation. For this purpose, we now take the factor $T(\omega,\Omega)$ as
\begin{align} 
T(\omega,\Omega)=Te^{-\frac{(\omega-\omega_{i0})^{2}}{2\beta^{2}}},
\end{align}
where the conversion rate has a maximum at $\omega_{i0}$ to achieve maximum conversion. It decreases significantly outside the range indicated by $\beta$. We update Eq. (11) and (14) to
\begin{align} 
g^{(2)}(t,t+\tau)=\frac{4T^{2}\beta^{2}\sigma_{P}\sigma_{-}}{\pi(2\beta^{2}+4\sigma_{P}^{2}+\sigma_{-}^{2})}e^{-\frac{\Omega^{2}+16\sigma_{P}^{2}\sigma_{-}^{2}(t+\tau-\tau_{0})^{2}+8\beta^{2}\sigma_{P}^{2}(2t+\tau-\tau_{0})^{2}+2\beta^{2}\sigma_{-}^{2}(\tau-\tau_{0})^{2}}{2(2\beta^{2}+4\sigma_{P}^{2}+\sigma_{-}^{2})}},
\end{align}
and
\begin{align} 
g^{(2)}(\tau)=\frac{T^{2}\beta^{2}\sigma_{-}}{\sqrt{2\pi(2\beta^{2}+4\sigma_{P}^{2}+\sigma_{-}^{2})(\sigma_{-}^{2}+2\beta^{2})}}e^{-\frac{\Omega^{2}}{2(2\beta^{2}+4\sigma_{P}^{2}+\sigma_{-}^{2})}}e^{-\frac{4(\tau-\tau_{0})^{2}\beta^{2}\sigma_{-}^{2}\sigma_{P}^{2}}{(2\beta^{2}+4\sigma_{P}^{2}+\sigma_{-}^{2})(\sigma_{-}^{2}+2\beta^{2})}}e^{-\frac{\beta^{2}\sigma_{-}^{2}(\tau-\tau_{0})^{2}}{(2\beta^{2}+4\sigma_{P}^{2}+\sigma_{-}^{2})}}\notag\\
\times\{erf[\sqrt{\frac{2(\sigma_{-}^{2}+2\beta^{2})}{2\beta^{2}+4\sigma_{P}^{2}+\sigma_{-}^{2}}}(2\tau-2\tau_{0}+T_{R})]-erf[\sqrt{\frac{2(\sigma_{-}^{2}+2\beta^{2})}{2\beta^{2}+4\sigma_{P}^{2}+\sigma_{-}^{2}}}(2\tau-2\tau_{0}-T_{R})]\}.
\end{align}
The expressions (19) and (20) reduce to equations (11) and (14) for $\beta^{2}\rightarrow\infty$. If the condition $T_{R}\gg(\tau-\tau_{0})$ is satisfied, then
\begin{align}
g^{(2)}(\tau)\propto e^{-\frac{\Omega^{2}}{2(2\beta^{2}+4\sigma_{P}^{2}+\sigma_{-}^{2})}}e^{-\frac{\beta^{2}\sigma_{-}^{2}}{(\sigma_{-}^{2}+2\beta^{2})}(\tau-\tau_{0})^{2}},
\end{align}
showing explicit dependence on the parameter $\beta$ and the frequency change $\Omega$.
%The FWHM depends both on $\sigma_-$ and $\beta$. As shown in Fig. 5(b), the ideal condition $\beta^{2}\rightarrow\infty$ is not required for a relatively sharp peak. When $\beta^{2}\approx\sigma_{-}$ the FWHM is only increased to $\sqrt{3/2}$ times its value for $\beta^{2}\rightarrow\infty$. 

\section{Hong–Ou–Mandel Interference of the Upshifted Signal and Original Idler Biphotons}

\begin{figure}[tb]
\centering
\includegraphics[width=10cm]{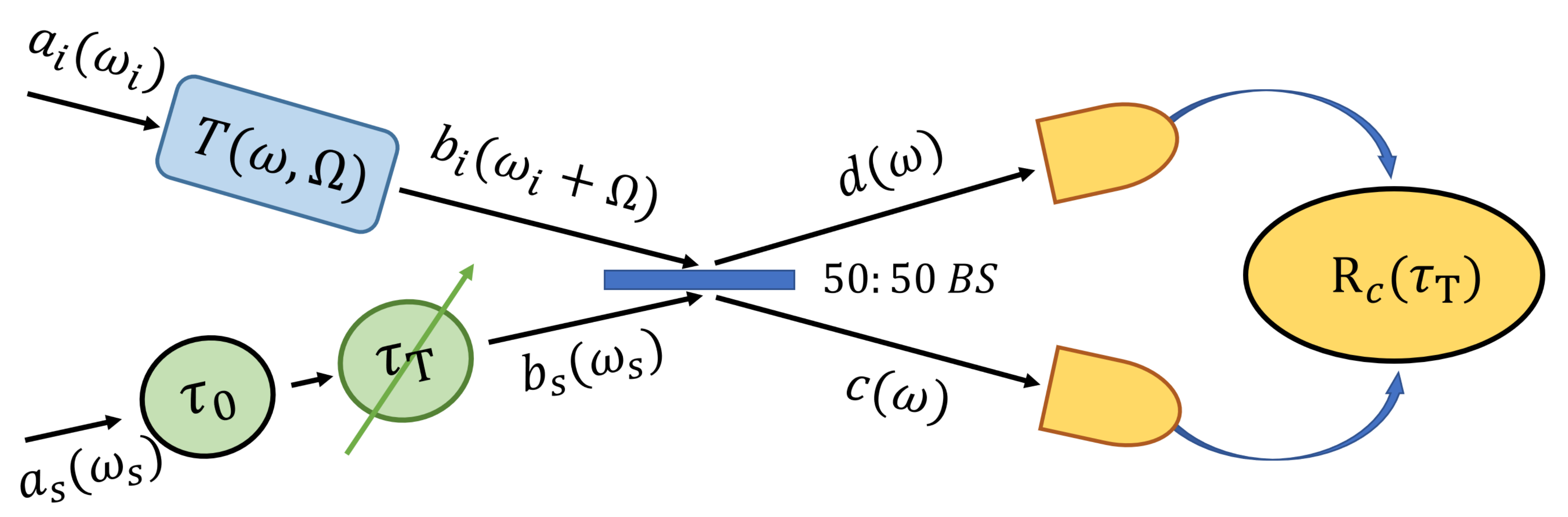}
\caption{Experimental setup for the Hong–Ou–Mandel (HOM) measurement of the upshifted idler and original signal biphotons, where $\tau_T$ is a tunable phase.}
\label{Fig4}
\end{figure}

The Hong–Ou–Mandel (HOM) interferometer \cite{hong1987measurement, bouchard2020two, hochrainer2022quantum}, a widely used tool in measuring biphoton joint frequency distribution, especially for time scales of 10's of fs \cite{lyons2018attosecond, PhysRevApplied.1.034004, ndagano2022quantum, eshun2021investigations}. We, therefore, study the two-photon interference between the up-converted idler and the signal photon. By introducing a tunable phase $\tau_T$ in the path of the signal photon, as depicted in Fig. 3, it is possible to measure both the frequency shift $\Omega$ and phase shift $\tau_0$ in a single setup. For the signal photon, 
\begin{figure}[tb]
\centering
\includegraphics[width=10cm]{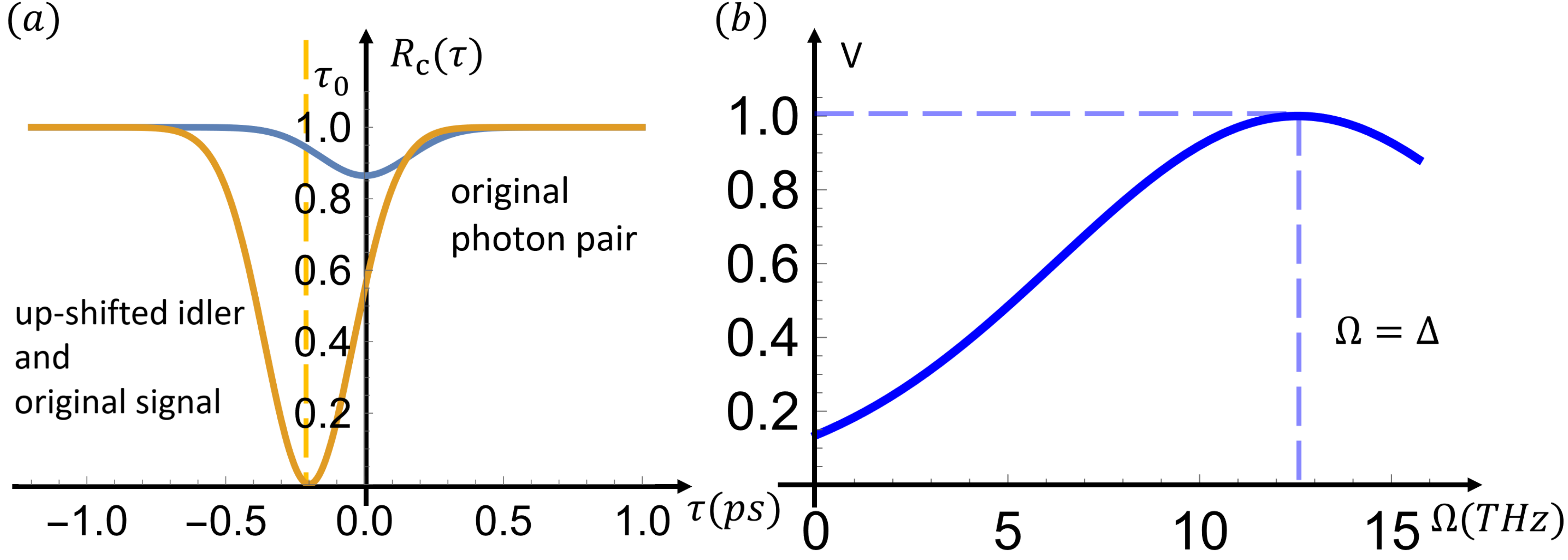}
\caption{(a) Normalized $R_c(\tau_T)$ function of the up-shifted idler and original signal photons (yellow), compared to the $R_c(\tau_T)$ of the original photon pair (blue), for $\sigma_{-} = 2\pi\times 1\:THz$, $\tau_0 = 0.2\:ps$, $\triangle = 2\pi\times 2\:THz$ and $\triangle-\Omega = 2\pi\times 0.05\:THz$. (b) HOM dip visibility as a function of the shifted frequency $\Omega$ when $\sigma_{-} = 2\pi\times 1\:THz$ and $\triangle = 2\pi\times 2\:THz$.}
\label{Fig5}
\end{figure}
\begin{align}
b_{s}(\omega)=a_{s}(\omega)e^{-i\omega(\tau_T+\tau_0)},
\end{align}
while the idler photon travels through the same up-conversion medium. After frequency conversion, the two photons are recombined at a 50:50 beam-splitter. The output of the beam-splitter is expressed as
\begin{align}
c(\omega)=\frac{b_{i}(\omega)+ib_{s}(\omega)}{\sqrt{2}}, \notag\\
d(\omega)=\frac{ib_{i}(\omega)+b_{s}(\omega)}{\sqrt{2}}.
\end{align}
As a result, the averaged coincidence rate of the output fields
\begin{align}
R_{c}(\tau_T)=\frac{1}{4\pi^2}\int dt\int d\tau\expval{d^{\dagger}(t)c^{\dagger}(t+\tau)c(t+\tau)d(t)}
\end{align}
is measured, where we assume the detection time is much larger than the pump correlation time $1/\sigma_p$ and the entanglement time $1/\sigma_-$. Changing the modes to the frequency domain $d(t)=\frac{1}{\sqrt{2\pi}}\int d\omega_{1}d(\omega_{1})e^{-i\omega_{1}t}$ and $c(t+\tau)=\frac{1}{\sqrt{2\pi}}\int d\omega_{2}c(\omega_{2})e^{-i\omega_{2}(t+\tau)}$, we obtain
\begin{align}
R_{c}(\tau_T)=\frac{1}{8\pi^2}&[\int\int d\omega_{1}d\omega_{2}f^{*}(\omega_{1}-\Omega,\omega_{2})f(\omega_{1}-\Omega,\omega_{2})T^{*}(\omega_{1},\Omega)T(\omega_{1},\Omega) \notag\\
-\int\int& d\omega_{1}d\omega_{2}f^{*}(\omega_{1}-\Omega,\omega_{2})f(\omega_{2}-\Omega,\omega_{1})T^{*}(\omega_{1},\Omega)T(\omega_{2},\Omega)e^{-i(\omega_{1}-\omega_{2})(\tau_T+\tau_0)}]. \notag\\
\end{align}
For a system with the same constant conversion rate assumption and photon distribution as in Eq. (2), we obtain
\begin{align}
R_{c}(\tau_T)=\frac{T^{2}}{8\pi^{2}}(1-e^{-\frac{(\Omega-\triangle)^{2}}{2\sigma_{-}^{2}}}e^{-\frac{1}{2}\sigma_{-}^{2}(\tau_T+\tau_0)^{2}}).
\end{align}
The FWHM of $R_{c}(\tau_T)$ is the same as the simple $g^{(2)}$ function as in Eq. (15), (16). The position of the peak in the HOM dip provides information on the phase difference, while the visibility $e^{-\frac{(\Omega-\triangle)^{2}}{2\sigma_{-}^{2}}}$ of the dip reveals the frequency shift $\Omega$. 
As shown in Fig. 4, the visibility of the HOM dip increases if the frequency difference between the signal and idler photons decreases due to the frequency conversion. 
Note that Eq. (26) shows that the visibility of the HOM dip remains sufficient as long as the parameter $f=\frac{\Omega-\triangle}{\sigma_{-}} \lesssim 1$, which is the case for reported experiments \cite{fan2016integrated, chen2021single, zhu2022spectral}.
%For example, $f=0.2$ for the parameters of reference \cite{fan2016integrated}, and nearly zero for the parameters in reference \cite{chen2021single} and \cite{zhu2022spectral}. 
%This is particularly evident when the original frequency difference is large, resulting in an almost invisible $R_c(\tau_T)$ dip. It is noteworthy to mention that the coincidence resolution time, as stated in the experiment conducted in \cite{tyumenev2022tunable}, is 80 ps, which leads to a highly restricted $\sigma_-$ value, demanding an incredibly precise frequency match after undergoing the frequency conversion process, in order to be compatible with the Hong–Ou–Mandel (HOM) interferometer. However, as long as the magnitude of $\Omega-\triangle$ is similar or even smaller than $\sigma_{-}$, the visibility of $R_c(\tau_T)$ becomes visible. This is demonstrated in the experimental results in \cite{fan2016integrated, chen2021single, zhu2022spectral}, and is shown in Fig. 4. 

\begin{figure}[tb]
\centering
\includegraphics[width=10cm]{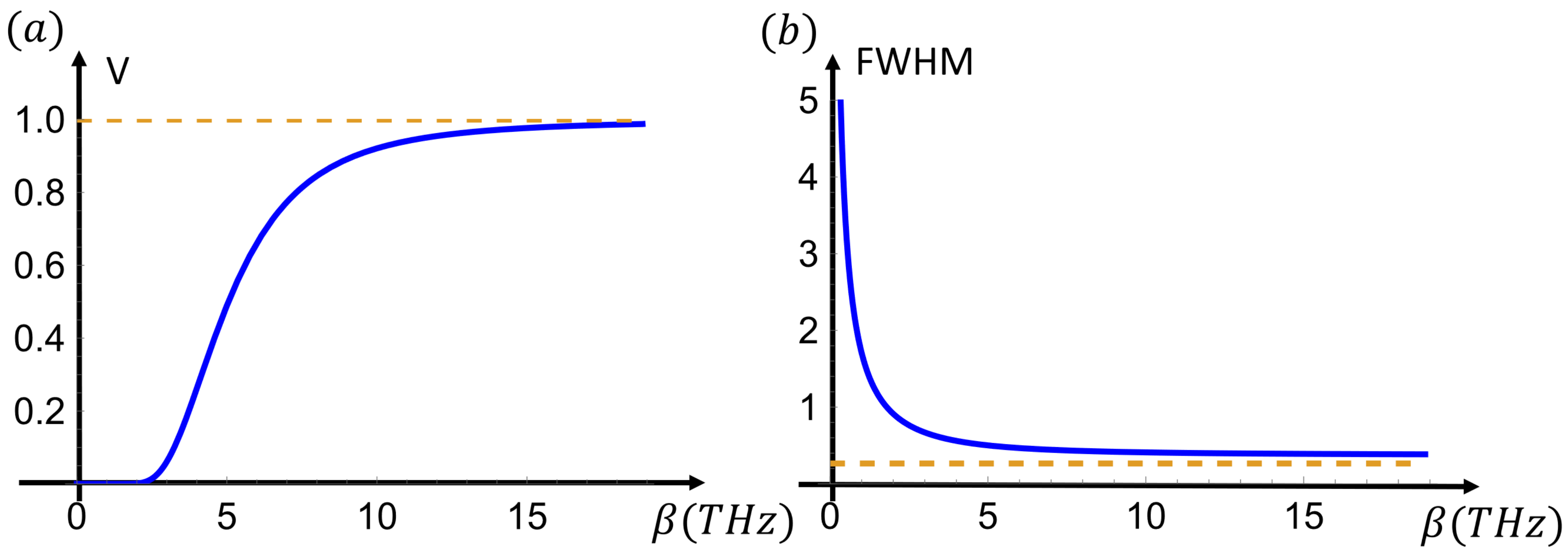}
\caption{(a) HOM dip visibility, and (b) FWHM for both $R_c(\tau_T)$ and $g^{{2}}(\tau)$, as a function of the phase-matching factor $\beta$ when $\sigma_{-} = 2\pi\times 1\:THz$, $\sigma_{p} = \sigma_{-}/10$, $\triangle = 2\pi\times 2\:THz$ and $\triangle-\Omega = 2\pi\times 0.05\:THz$. The dashed lines indicate the limits when $\beta\gg\sigma_{-}$.}
\label{Fig6}
\end{figure}

Next, we examine the effects of imperfect phase-matching in up-conversion using Eq. (18) and Eq. (25); we obtain
\begin{align}
R_{c}(\tau_{T})&=\frac{\sqrt{2}T^{2}\beta}{8\pi^{2}}[\frac{1}{\sqrt{4\sigma_{P}^{2}+\sigma_{-}^{2}+2\beta^{2}}}e^{-\frac{2\Omega^{2}}{4\sigma_{P}^{2}+\sigma_{-}^{2}+2\beta^{2}}} \notag\\
&-\frac{\beta}{\sqrt{(\beta^{2}+2\sigma_{P}^{2})(2\beta^{2}+\sigma_{-}^{2})}}e^{-\frac{(\Omega+\triangle)^{2}}{4(\beta^{2}+2\sigma_{P}^{2})}}e^{-\frac{(\Omega-\triangle)^{2}}{2\sigma_{-}^{2}}}e^{-\frac{\sigma_{-}^{2}\beta^{2}(\tau_{T}+\tau_{0})^{2}}{2\beta^{2}+\sigma_{-}^{2}}}],
\end{align}
which reduces to Eq. (26) for $\beta^{2}\rightarrow\infty$. Note that Eq. (27) shares the same FWHM as Eq. (21). From Eq. (27), we obtain the visibility of the HOM dip
\begin{align}
V=\frac{\beta\sqrt{4\sigma_{P}^{2}+\sigma_{-}^{2}+2\beta^{2}}}{\sqrt{(\beta^{2}+2\sigma_{P}^{2})(2\beta^{2}+\sigma_{-}^{2})}}e^{-\frac{(\Omega+\triangle)^{2}}{4(\beta^{2}+2\sigma_{P}^{2})}}e^{-\frac{(\Omega-\triangle)^{2}}{2\sigma_{-}^{2}}}e^{\frac{2\Omega^{2}}{4\sigma_{P}^{2}+\sigma_{-}^{2}+2\beta^{2}}}.
\end{align}
The visibility is maximized when $\Omega=\triangle$ as
\begin{align}
V=\frac{\beta\sqrt{4\sigma_{P}^{2}+\sigma_{-}^{2}+2\beta^{2}}}{\sqrt{(\beta^{2}+2\sigma_{P}^{2})(2\beta^{2}+\sigma_{-}^{2})}}e^{-\frac{(8\sigma_{P}^{2}+\sigma_{-}^{2}+4\beta^{2})\Omega^{2}}{(4\sigma_{P}^{2}+\sigma_{-}^{2}+2\beta^{2})(\beta^{2}+2\sigma_{P}^{2})}}.
\end{align}
When $\beta\gg\sigma_{-}$ and $\sigma_{P}$, the visibility $V\rightarrow 1$. However, a very satisfactory visibility of 0.96 can be obtained when $\beta\sim 2\sigma_{-}$ for a system as shown in Fig. 5(a). The FWHM depends both on $\sigma_-$ and $\beta$. As shown in Fig. 5(b), when $\beta\sim 2\sigma_{-}$ the peak is only expanded by 50\% compared to the value when $\beta\gg\sigma_{-}$. 
%The analysis thus shows that the HOM dip is not very sensitive to the precise phase-matching condition.
%By measuring the visibility, one is thus able to measure both the average frequency shift $\Omega$ and the variance $2\beta^2$, which indicates the range of frequency that certain materials' ability for up-conversion can cover. In this way, the HOM setup not only provides a powerful tool to quantify the performance of the up-conversion process but also gives insights into the properties of the medium used. Moreover, the measurement of the visibility and FWHM of the HOM dip can be easily performed through various experimental setups, making it an accessible and convenient method for the characterization of the up-conversion process. 

\section{Conclusions}

The results of our theoretical study provide a comprehensive insight
into the effect of frequency up-conversion on the time-dependent quantum
correlation of a photon pair. By deriving the second-order
correlation function, we demonstrate that the full width at
half maximum (FWHM) remains unchanged, while the peak
height and position shift after the up-conversion procedure.
This finding agrees with the recent experiment \cite{tyumenev2022tunable}, where the correlation time was within reach of the
detector resolution time. 
This is true if the phase matching is satisfied over the spectral width of the idler photon.
To gain a deeper understanding of the impact of the frequency up-conversion
on the quantum correlation, we examined the two-photon Hong-Ou-Mandel interferometry. We showed
how the visibility of the two-photon interference was sensitive to the magnitude of the frequency change in the up-conversion process, the bandwidth of the signal photon, and the phase matching factor $\beta$.
These results are fairly general and are expected to apply to other types of frequency conversion.

\begin{backmatter}

\section*{Funding}
JW and GSA thank the support of Air Force Office of Scientific Research (Award N%
\textsuperscript{\underline{o}} FA-9550-20-1-0366) and the Robert A Welch
Foundation (A-1943-20210327). 
AVS thanks Welch Foundation (Grant N%
\textsuperscript{\underline{o}} A-1547) for support.

\section*{Disclosures} The authors declare no conflicts of interest.

\section*{Data Availability} No data were generated or analyzed in the presented research.

\end{backmatter}

%%%%%%%%%%%%%%%%%%%%%%% References %%%%%%%%%%%%%%%%%%%%%%%%%

%%%%%%%%%% If using BibTeX:
\bibliography{sample}

%%%%%%%%%% If preparing manually:
% \begin{thebibliography}{1}
% \newcommand{\enquote}[1]{``#1''}

% \bibitem{Zhang:14}
% Y.~Zhang, S.~Qiao, L.~Sun, Q.~W. Shi, W.~Huang, L.~Li, and Z.~Yang,
%   \enquote{Photoinduced active terahertz metamaterials with nanostructured
%   vanadium dioxide film deposited by sol-gel method,}
%   {\protect\JournalTitle{Optics Express}} \textbf{22}, 11070--11078 (2014).

% \bibitem{Optica}
% {Optica}, \enquote{{Optica Publishing Group},}
%   \url{http://www.opg.optica.org}.

% \bibitem{FORSTER2007}
% P.~Forster, V.~Ramaswamy, P.~Artaxo, T.~Bernsten, R.~Betts, D.~Fahey,
%   J.~Haywood, J.~Lean, D.~Lowe, G.~Myhre, J.~Nganga, R.~Prinn, G.~Raga,
%   M.~Schulz, and R.~V. Dorland, \enquote{Changes in atmospheric consituents and
%   in radiative forcing,} in \enquote{Climate Change 2007: The Physical Science
%   Basis. Contribution of Working Group 1 to the Fourth assesment report of
%   Intergovernmental Panel on Climate Change,}  S.~Solomon, D.~Qin, M.~Manning,
%   Z.~Chen, M.~Marquis, K.~B. Averyt, M.~Tignor, and H.~L. Miler, eds.
%   (Cambridge University Press, 2007).

% \end{thebibliography}

\end{document}